\documentclass[12pt]{article}  
\usepackage{graphicx}
\usepackage{colordvi}
\usepackage{color}
\begin{document}

\title{The dynamics of the early universe in a model with radiation and a generalized Chaplygin gas}

\author{G. A. Monerat and C. G. M. Santos \\ 
Departamento de Modelagem Computacional, \\
Instituto Polit\'{e}cnico, \\
Universidade do Estado do Rio de Janeiro, \\
CEP 28625-570, Nova Friburgo, RJ, Brazil.\\
monerat@uerj.br, cassia.gmsantos@gmail.com
\and G. Oliveira-Neto\\
Departamento de F\'{\i}sica, \\
Instituto de Ci\^{e}ncias Exatas, \\ 
Universidade Federal de Juiz de Fora,\\
CEP 36036-330, Juiz de Fora, MG, Brazil.\\
gilneto@fisica.ufjf.br
\and E. V. Corr\^{e}a Silva and L. G. Ferreira Filho \\ 
Departamento de Matem\'{a}tica F\'{\i}sica e Computa\c{c}\~{a}o,\\
Faculdade de Tecnologia, \\
Universidade do Estado do Rio de Janeiro,\\
Rodovia Presidente Dutra Km 298, P\'{o}lo Industrial, \\
CEP 27537-000, Resende, RJ, Brazil.\\
eduardo.vasquez@pq.cnpq.br, gonzaga@uerj.br}

\maketitle

\abstract{The early universe is modeled through the quantization of a
Friedmann-Robertson-Walker model with positive curvature of the spatial hypersurfaces. In this model, the 
universe is filled by two fluids: radiation and a generalized Chaplygin gas. The quantization 
of this model is made following the prescriptions due to J. A. Wheeler and B. DeWitt. 
Using the Schutz's formalism, the time notion is recovered and the Wheeler-DeWitt 
equation transforms into a time dependent Schr\"{o}dinger equation, which rules 
the dynamics of the early universe, under the action of an effective potential 
$V_{eff}$. That potential, depends on three parameters. Depending on the values of these parameters, $V_{eff}$ may have two different shapes. $V_{eff}(a)$ may have the shape of a barrier or the shape of a well followed by a barrier. We solve, numerically, the appropriate time dependent Schr\"{o}dinger equation and obtain the time evolution 
of an initial wave function, for both cases. These wave functions satisfy suitable boundary 
conditions. For both shapes of $V_{eff}$, we compute the tunneling probability, which is a function of
the mean kinetic energy associated to the radiation energy $E_m$ and of the three parameters of the generalized 
Chaplygin gas: $\alpha$, $A$ and $B$. The tunneling probabilities, for both shapes of $V_{eff}$, indicates 
that the universe should nucleate with the highest possible values of $E_m$, $\alpha$, $A$ and $B$.
Finally, we study the classical universe evolution after the wavefunction has tunneled $V_{eff}$.
The calculations show that the universe may emerge from the Planck era in an inflationary phase.}

%\begin{keyword}

%Quantum cosmology \sep Wheeler-DeWitt equation \sep Generalized Chaplygin gas \sep Tunneling probability

%\PACS 04.40.Nr \sep 04.60.Ds \sep 98.80.Qc

%\end{keyword}

\section{Introduction}

The first attempt to obtain a quantum mechanical model of a cosmological spacetime was made by
Bryce DeWitt in 1967 \cite{dewitt}. He used the canonical quantization method applied to
general relativity (GR) and specialized to a Friedmann-Robertson-Walker (FRW) spacetime, with positive
three-dimensional spacelike hypersufaces. The matter content of the cosmological model was a
perfect fluid of dust. After that initial work many others followed. One of the main motivations 
for the works, in that area, is the elimination of the so-called {\it Big Bang} singularity.
That singularity was shown to be present, under very general and reasonable conditions,
in many cosmological solutions to the general relativity equations \cite{hawking}. In 1983,
J. B. Hartle and S. W. Hawking gave an important contribution to that area \cite{hawking1}.
They proposed to quantize general relativity using the path integral approach. They exemplified
their proposal quantizing a cosmological FRW spacetime, with positive three-dimensional spacelike 
hypersufaces. The matter content of the cosmological model was a conformal scalar field. They 
computed the ground state wavefuncion for that model and showed, how, our Universe, could had emerged 
from a singularity free Euclidean universe, through a quantum mechanical tunneling process. There were
other important contributions to those ideas concerning the creation of the Universe from nothing 
\cite{vilenkin,vilenkin1}.

In 1998, important observations showed that our Universe is expanding in an accelerated rate \cite{riess0,perlmutter}.
Those observations changed in a dramatic way our understanding of the Universe. One possible way to
interpret those observations is to consider the existence of an `exotic' type of matter in our Universe.
Among the unusual properties of that `exotic' matter, one may mention that it must be repulsive, instead of
attractive as the ordinary matter. That property implies that if one has a certain amount of that `exotic' matter, 
in a finite region, it produces a negative pressure in that region. That explanation, for the present
accelerated expansion of the Universe, has been pursued by many physicists. They have proposed many
different types of `exotic' matter, which are presently known as dark energy. Two important examples
of dark energy is the Chaplygin Gas \cite{chaplygin,moschella} (CG) and its generalization the generalized Chaplygin gas \cite{moschella,bilic,bento} (GCG).
The energy density of the GCG depends on three parameters $A$, $B$ and $\alpha$. When one sets $\alpha=1$, one recovers the CG. 
Since their introductions, in the literature of cosmology, several important works have been produced, studying quantum aspects 
of both gases CG and GCG \cite{paulo,gil,pedram0,pedram,majumder,pedram1,pedram2}.

In the present paper, we study the birth of a Friedman-Robertson-Walker (FRW) universe, with positive spatial 
sections, based on GR, coupled to a radiation perfect fluid and a generalized Chaplygin gas, due to a quantum tunneling 
process. The two fluids present in the model try to describe different stages
of the matter content of our Universe. Initially, the matter content of the Universe is better described by 
radiation, then, at later stages, by a GCG. We consider the canonical quantization of the model. Therefore, we start 
using the ADM formalism to write the gravitational sector of the hamiltonian \cite{ADM}. Then, we write the radiation 
fluid hamiltonian using the Schutz variational formalism \cite{schutz,schutz1}. We do that in order to obtain a time 
variable, at the quantum mechanical level, which is associated to a radiation fluid degree of freedom. The Chaplygin 
gas is introduced in the Lagrangian of the model through its matter density, following the literature of the area 
\cite{hawking,nasseri,paulo,gil}. We canonically quantize the model using the Dirac's formalism for constrained systems
\cite{dirac,dirac1,dirac2}. As the result of that quantization process, we obtain a Wheeler-DeWitt equation in the
form of a time dependent Schr\"{o}dinger equation, with an effective potential $V_{eff}$. We notice that, depending on 
the values of the GCG parameters, $V_{eff}$ may have two distinct shapes. The first one, is the shape of 
a barrier and the second one is the shape of a well followed by a barrier. Quantum mechanically, the universe may tunnel
through the potential barriers, present in both shapes of $V_{eff}$. Here, we compute the tunneling probability ($TP$), for both
shapes of $V_{eff}$. In order to do that, we solve numerically the appropriate time dependent Schr\"{o}dinger equation and obtain the time evolution of an initial wave function, for both shapes of $V_{eff}$. These wave functions satisfy suitable
boundary conditions. For both shapes of $V_{eff}$, we compute the tunneling probability, which is a function of
$E_m$ and of the three parameters of the generalized Chaplygin gas: $\alpha$, $A$ and $B$.
If we take into account that those parameters are related to: (i) the cosmological constant, for great values of the scale factor ($A$); (ii) the constant present in the energy density of a dust perfect fluid, for small values of the scale factor ($B$); (iii) the type of generalized Chaplygin gas ($\alpha$); (iv) the mean kinetic energy associated to the radiation energy ($E_m$); then, from the tunneling probabilities, we should gain information on the most probable values of the cosmological constant ($\Lambda$), the energy density of that dust perfect fluid, the radiation energy and the type of generalized Chaplygin gas. Finally, from the appropriate tunneling wavefunction, with a suitable approximation, we obtain the values of the scale factor and its first derivative with respect to the conformal time, just after the universe has emerged from the right of the barrier. Using these initial conditions and the classical dynamical equations, we compute the classical evolution of the scale factor and show that it leads to an universe that expand in an inflationary way.

In fact, the quantum cosmology of a FRW model, with positive spatial sections, based on GR and coupled to a GCG, 
have already been studied in the literature \cite{paulo}. In fact, there are several differences between
the present paper and Ref. \cite{paulo}. Here, besides the GCG, we consider a radiation perfect fluid in the matter content
of the model. Therefore, due to the presence of that radiation fluid and the Schutz variational formalism, at the quantum level, 
we obtain a time dependent Schr\"{o}dinger equation. In Ref. \cite{paulo}, they consider only the GCG and obtain a time 
independent Wheeler-DeWitt equation. Here, we find the wave function of the Universe by solving exactly, using a numerical
technique, the time dependent Schr\"{o}dinger equation. In Ref. \cite{paulo}, the authors use a WKB approximation to solve
the time independent Wheeler-DeWitt equation, for an approximate potential that describes the GCG, under certain conditions.
They consider, only the case where that approximated potential has the shape of a well followed by a barrier.
It is important to mention that we have, also, few qualitative agreements between both papers. The most important agreement 
concerns the $TP$, computed in both papers. Their approximated, analytical expression for $TP$ as a function 
of $A$, $B$ and $\alpha$, grows with the increase of these parameters, for the boundary conditions of the tunneling wave-function 
\cite{vilenkin}. Those results are in agreement with our numerical results for $TP$ as a function of the same parameters. 
It is important to mention that the present paper may be considered an extension of a previous paper \cite{gil}, for the most
general case of a GCG. There, some of us have studied the quantum cosmology model based on GR, of a similar model where,
instead of a GCG, in the matter content of the model, we had a CG. 

This paper is organized as follows. In next section, the classical model is presented and the effective potential
$V_{eff}(a)$ is introduced. In section III, 
the classical model is quantized. In section IV, we study models where $V_{eff}(a)$ has a barrier shape.  The wave function describing 
the quantum system is calculated and the tunneling probability as a function of $E_m$, $A$, $B$ and $\alpha$ is computed, for 
that case. In section V, we study models where $V_{eff}(a)$ has a well followed by a barrier shape. The wave function describing 
the quantum system is calculated and the tunneling probability as a function of $E_m$, $A$, $B$ and $\alpha$ is computed, for 
that case. In section VI, we examine how the classical universe appears after the tunneling process and gives rise to an
inflationary phase for both shapes of $V_{eff}$, described in the previous sections. The conclusions are presented in section VII.

\section{The Classical Model}

The FRW cosmological models are characterized by the
scale factor $a(t)$ and have the following line element,
\begin{equation}  \label{1}
ds^2 = - N^2(t) dt^2 + a^2(t)\left( \frac{dr^2}{1 - kr^2} + r^2 d\Omega^2\right) ,
\end{equation}
where $d\Omega^2$ is the line element of the two-dimensional sphere with
unitary radius, $N(t)$ is the lapse function and $k$ gives the type of
constant curvature of the spatial sections. Here, we are considering the 
case with positive curvature $k=1$ and we are using the natural
unit system, where $\hbar=c=8\pi G=1$. The matter content of the model is
represented by a GCG and a radiation perfect fluid.

The generalized Chaplygin gas is defined through the equation of state \cite{bilic,bento},
\begin{equation}
p_c = - \frac{\bar{A}}{\rho_c^\alpha} ,
\label{2}
\end{equation}
where $\bar{A}$ is a positive constant, $0 < \alpha \leq 1$ and $p_c$ and $\rho_c$ are the generalized Chaplygin gas
pressure and energy density, respectively. The standard Chaplygin gas \cite{chaplygin,moschella} corresponds to $\alpha = 1$.
On the other hand, the radiation perfect fluid has a different equation of state, given by,
\begin{equation}
p_r = \frac{1}{3}{\rho_r} ,
\label{3}
\end{equation}
where $p_r$ and $\rho_r$ are the radiation fluid pressure and energy density, respectively.

Our starting point, in order to describe the dynamics of the model, is the Lagrangian of the model ($L$), including the GR and matter Lagrangians,
\begin{equation}
L = L_G + L_{GCG} + L_R ,
\label{4}
\end{equation}
where $L_G = \sqrt{-g}R/2$ is the GR Lagrangian, $R$ is the Ricci scalar, $g$ is the determinant of the metric, $L_{GCG}$ is the 
generalized Chaplygin gas Lagrangian and $L_R$ is the radiation fluid Lagrangian. 
 
Using the metric Eq. (\ref{1}), we may rewrite the Lagrangian $L$ Eq. (\ref{4}), in the following way,
\begin{equation}
L = L_G + L_{GCG} + L_R = - 3\left(\frac{\dot a^2 a}{N} - Na\right) - a^3 N\rho_c + a^3 N p_r ,
\label{5}
\end{equation}
where the dot means the derivative of $a$ with respect to time. Now, in order to obtain an explicit expression of $\rho_c$, as a function of $a$, we use
the conservation law for the stress-energy tensor of the GCG and its equation of state Eq. (\ref{2}). After some calculation, we obtain,
\begin{equation}
\rho_c(a) = \left(\bar{A} + \frac{\bar{B}}{a^{3(1+\alpha)}}\right)^{1/(1+\alpha)} ,
\label{6}
\end{equation}
where $\bar{B}$ is a constant that may be positive or negative. 
It is interesting to notice that $\rho_c(a)$ Eq. (\ref{6}) has two important asymptotic limits. When, $a \to 0$ and
$\bar{B}>0$, it behaves as the energy density of a dust perfect fluid. Therefore, in that situation, $\bar{B}$ is related to the energy density of the dust perfect fluid, at a certain chosen time. On the other hand, when, $a \to \infty$, it behaves as
a constant ($(\bar{A})^{1/(1+\alpha)}$) which may be associated to a cosmological constant.

Here, we treat the radiation perfect fluid using the Schutz variational formalism \cite{schutz,schutz1}. As we mentioned, above, the
advantage of that formalism is the introduction of a time variable associated to the perfect fluid degree of freedom. We start by writing
the perfect fluid four-velocity in terms of six thermodynamical potentials,
\begin{equation}
U_\nu = \frac{1}{\mu}(\epsilon_{,\nu} + \zeta\beta_{,\nu} + \theta
S_{,\nu}),
\label{7}
\end{equation}
where $U_\nu$ is the perfect fluid four-velocity and $\mu$, $\epsilon$, $\zeta$, $\beta$, $\theta$ and $S$ are the six thermodynamical potentials.
The potentials are identified as: $\mu$ is the specific enthalpy, $S$ is the specific entropy, $\zeta$ and $\beta$ are connected with rotation,
$\epsilon$ and $\theta$ have no clear physical meaning. In the present model the potentials $\zeta$ and $\beta$ are zero because they do not
contribute in FRW cosmological space-times.

Now, if one applies the normalization condition: $U^\nu U_\nu = - 1$, to Eq. (\ref{7}), one obtains
the specific enthalpy $\mu$ as a function of the other thermodynamical potentials. With the exception
of $\zeta$ and $\beta$, which will not contribute in the present model, as explained above. That
equation, for $\mu$, may be given in the following way,
\begin{equation}
\label{8}
\mu = \left( -g^{00}(\epsilon_{, 0} +\theta S_{, 0})^2 \right)^{\frac{1}{2}} = \frac{\dot{\epsilon} + \theta \dot{S}}{N}
\end{equation}
Now, using the equation of state for the radiation fluid Eq. (\ref{3}), Eq. (\ref{8}) and the following equation which relates $\rho$,
$\mu$ and $S$, derived with the aid of the first law of thermodynamics \cite{schutz,schutz1},
\begin{equation}
\label{8,5}
\rho = \left(\frac{3\mu}{4}\right)^4e^{-3S},
\end{equation}
we obtain the following expression for the radiation fluid Lagrangian Eq. (\ref{5}),
\begin{equation}
L_R = \frac{27}{256} N^{-3} a^3 (\dot\epsilon + \theta\dot S)^{4} e^{-3S}.
\label{9}
\end{equation}
Finally, we may write the following expression for $L$ Eq. (\ref{5}), with the aid of $\rho_c(a)$ Eq. (\ref{6}) and $L_R$ Eq. (\ref{9}),
\begin{equation}
\label{9,3}
L = - 3\left(\frac{\dot a^2 a}{N} - Na\right) - a^3 N\left(\bar{A} + \frac{\bar{B}}{a^{3(1+\alpha)}}\right)^{1/(1+\alpha)} + \frac{27}{256} N^{-3} a^3 (\dot\epsilon + \theta\dot S)^{4} e^{-3S}
\end{equation}
Now, in order to write Hamiltonian of the model, including the gravitational, the generalized Chaplygin gas and the radiation fluid sectors, 
we start computing the canonically conjugated momenta from $L$ Eq. (\ref{9,3}),
\begin{eqnarray}
\label{9,5}
P_a & = & \frac{\partial L}{\partial \dot{a}} = - \frac{6a\dot{a}}{N},\\
P_\epsilon & = & \frac{\partial L}{\partial \dot{\epsilon}} = \frac{27}{64} N^{-3} a^3 (\dot{\epsilon} + \theta\dot{S})^{3} e^{-3S}, \\
P_S & = & \frac{\partial L}{\partial \dot{S}} = \theta P_\epsilon .
\end{eqnarray}

We may write the Hamiltonian of the model ($H$), with the aid of the canonically conjugated momenta Eqs. (12), (13) and (14) and the Lagrangian $L$ Eq. (\ref{9,3}),
\begin{equation}
\label{10}
H = -\frac{NP_a^2}{12a} - 3Na + a^3 N\left(\bar{A} + \frac{\bar{B}}{a^{3(1+\alpha)}}\right)^{1/(1+\alpha)} + NP_\epsilon^{\frac{4}{3}}\frac{e^S}{a}.
\end{equation}
If one introduces the following canonical transformations,
\begin{equation}
T = -P_Se^{-S}P_\epsilon^{-\frac{4}{3}} \quad , \quad P_T =
P_\epsilon^{\frac{4}{3}}e^S \quad , \quad \bar\epsilon = \epsilon -
(\frac{4}{3})\frac{P_S}{P_\epsilon} \quad , \quad \bar P_\epsilon =
P_\epsilon \quad ,
\label{11}
\end{equation}
which generalizes the ones used in \cite{rubakov}, the Hamiltonian of the radiation fluid $H_R = NP_\epsilon^{\frac{4}{3}}\frac{e^S}{a}$ (\ref{10}) may be written in a simpler way,
\begin{equation}
H = -\frac{NP_a^2}{12a} - 3Na + a^3 N\left(\bar{A} + \frac{\bar{B}}{a^{3(1+\alpha)}}\right)^{1/(1+\alpha)} + \frac{NP_T}{a}.
\label{12}
\end{equation}
Observing $H$ Eq. (\ref{12}), one notices that the only remaining canonical variable associated to
the radiation fluid is $P_T$. Since it appears linearly in that equation, at the quantum level, it
may be identified with a time variable. 

Finally, the homogeneous and isotropic FRW model with positive
curvature of the spatial sections, radiation and generalized Chaplygin gas,
may be represented by a Hamiltonian with two degrees of freedom in
the form
\begin{equation}
H = \frac{1}{12}P_{a}^2 + V_{eff}(a) - P_{T}, 
\label{sec1eq1}
\end{equation}
where we choose the gauge where $N = a$, $P_{a}$ and $P_{T}$ are, respectively, the momenta
canonically conjugated to the scale factor and the variable that describes the perfect fluid, 
and $V_{eff}(a)$ is the effective potential given by,
\begin{equation}
V_{eff}(a) = 3a^2 - a^4\left(\frac{1}{\pi}\right)^{\frac{2}{(1+\alpha)}}\left(A + 
\frac{B}{a^{3+3\alpha}}\right)^{\frac{1}{(1+\alpha)}},
\label{sec1eq2}
\end{equation}
where $A=\pi^2\bar{A}$ and $B=\pi^2\bar{B}$. The quantum cosmology version of the standard Chaplygin gas model was treated in \cite{gil}.

\begin{figure}[h!]
\begin{center}
\includegraphics[height=5.0cm,width=6.0cm]{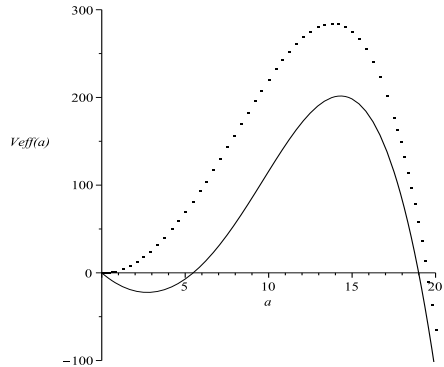}
\end{center}
\caption{$V_{eff}(a)$ for the barrier case: (i) $A=0.001,\, B=0.001$ (dotted line); and for the well followed by the barrier case: (ii) $A=0.001,\, B=2000$ (solid line). For both cases we use $\alpha=0.9$.}
\label{fpotenciais}
\end{figure}

Studying the expression of $V_{eff}(a)$ (\ref{sec1eq2}), we notice that, it may have three, two or one root, depending on the values of the parameters $A$ and $B$, for fixed $\alpha$. 
One of the roots will always be $a=0$. Here, we are going to restrict our attention to the case where $V_{eff}(a)$ (\ref{sec1eq2}) has three roots. In that case, if we increase the value of $a$ from
$a=0$, there will be the formation of a well, which minimum value is always negative, followed by a barrier, which maximum value is always positive. We are going to consider two different situations.
In the first one, the minimum value of the well is very close to zero and the maximum value of the barrier is significantly greater than zero. Therefore, the potential behaves as if there is just a barrier. 
In the second case, 
the minimum value of the well is significantly smaller than zero and, again, the maximum value of the barrier is significantly greater than zero. Therefore, the potential behaves as if there is
a well followed by a barrier. We want to study the tunneling probability in those two situations and compare the results between them. Examples, one for each case, are shown in Figure \ref{fpotenciais}.
Finally, for large values of $a$, the potential behaves as if it was generated by a positive cosmological constant. 

\subsection{The dependence of $V_{eff}(a)$ with the parameters}

\subsubsection{The barrier case}

\paragraph{Dependence with $\alpha$} If we fix the values of $A$ and $B$, we notice that when $\alpha$ decreases the maximum value of $V_{eff}(a)$ increases, the value of $a$, that gives that maximum, moves to higher values of $a$ and the width of $V_{eff}(a)$ increases, for a fixed energy $E$. 

\paragraph{Dependence with $A$} If we fix the values of $\alpha$ and $B$, we notice that when $A$ decreases the maximum value of $V_{eff}(a)$ increases, the value of $a$, that gives that maximum, moves to higher values of $a$ and the width of $V_{eff}(a)$ increases, for a fixed energy $E$. 

\paragraph{Dependence with $B$} If we fix the values of $\alpha$ and $A$, we notice that when $B$ decreases the maximum value of $V_{eff}(a)$ increases, the value of $a$, that gives that maximum, moves to smaller values of $a$ and the width of $V_{eff}(a)$ increases, for a fixed energy $E$. 

\subsubsection{The well followed by a barrier case}

\paragraph{Dependence with $\alpha$} If we fix the values of $A$ and $B$, we notice that the maximum of $V_{eff}(a)$ behaves differently for two different regions of $\alpha$. In the first one, for $0.13 < \alpha \leq 1$, when $\alpha$ increases the maximum of $V_{eff}(a)$ decreases. On the other hand, in the second region, for $0 < \alpha \leq 0.13$, when $\alpha$ increases the maximum of $V_{eff}(a)$ increases. The value of $a$ that gives the maximum of $V_{eff}(a)$ moves to smaller values of $a$ when $\alpha$ increases, for all values of $\alpha$.

\noindent If we fix the values of $A$ and $B$, we notice that when $\alpha$ increases the minimum of $V_{eff}(a)$ increases, but it is always negative. The value of $a$ that gives the minimum of $V_{eff}(a)$ moves to the singularity at $a=0$, when $\alpha$ increases. Those two results are valid for all values of $\alpha$.

\noindent Now, for $0.13 < \alpha \leq 1$, if we fix the values of $A$, $B$ and an energy $E$, we notice that the width of the barrier in $V_{eff}(a)$ increases when $\alpha$ decreases. On the other hand, for $0 < \alpha \leq 0.13$, if we fix the values of $A$, $B$ and an energy $E$, we notice that the width of the barrier in $V_{eff}(a)$ decreases when $\alpha$ decreases. 
For fixed $A$, $B$ and $E$, the width of the well in $V_{eff}(a)$ increases when $\alpha$ decreases, for all values of $\alpha$.

\paragraph{Dependence with $A$}  If we fix the values of $\alpha$ and $B$, we notice that when $A$ decreases the maximum value of $V_{eff}(a)$ increases, the value of $a$, that gives that maximum, moves to higher values of $a$.
Those results are valid for all values of $\alpha$.

\noindent Now, for $0.13 < \alpha \leq 1$, if we vary $A$, keeping $\alpha$ and $B$ fixed, we notice that the minimum value of $V_{eff}(a)$, that is always negative, 
is not significantly modified. On the other hand, for $0 < \alpha \leq 0.13$, we notice that when $A$ decreases the minimum value of $V_{eff}(a)$ increases, but it is always negative. The value of $a$ that gives the minimum of $V_{eff}(a)$ moves to the singularity at $a=0$, when $A$ decreases.  

\noindent Now, if we fix the values of $\alpha$, $B$ and an energy $E$, we notice that the width of the barrier in $V_{eff}(a)$ increases when $A$ decreases, for all values of $\alpha$.
For $0.13 < \alpha \leq 1$, if we vary $A$, keeping $\alpha$ and $B$ fixed, we notice that the width of the well in $V_{eff}(a)$, is not significantly modified.
Finally, for $0 < \alpha \leq 0.13$, if we fix $\alpha$, $B$ and $E$, the width of the well in $V_{eff}(a)$ increases when $A$ increases.

\paragraph{Dependence with $B$} For all values of $\alpha$, if we fix the values of $\alpha$ and $A$, we notice that when $B$ decreases the maximum value of $V_{eff}(a)$ increases.
For $0.13 < \alpha \leq 1$, when $B$ decreases, the value of $a$, that gives the maximum of $V_{eff}(a)$, moves to smaller values of $a$. On the other 
hand, for $0 < \alpha \leq 0.13$, when $B$ decreases, the value of $a$, that gives the maximum of $V_{eff}(a)$, moves to higher values of $a$.

\noindent Now, for all values of $\alpha$, if we fix the values of $\alpha$ and $A$, we notice that when $B$ increases the minimum of $V_{eff}(a)$, that is always negative, decreases and the value of $a$, that gives the minimum, moves to higher values of $a$. 

\noindent Now, if we fix the values of $\alpha$, $A$ and $E$ and increase the value of $B$, the width of the barrier in $V_{eff}(a)$ decreases, for all values of $\alpha$. For fixed $\alpha$, $A$ and $E$, the width of the well in $V_{eff}(a)$ increases when $B$ increases, for all values of $\alpha$.

\subsection{The classical dynamics}

The classical dynamics of the model is governed by the Hamilton's equations. For the present model, it is given by the following non-linear autonomous first-order system,
\begin{equation}
\left\{
\begin{array}{lll}
\dot{a} = + \frac{\partial H}{\partial P_{a}} &=& \frac{1}{6}P_{a} \\
\dot{P_{a}} = - \frac{\partial H}{\partial a} &=& -6a+4a^3\left(\frac{1}{\pi}\right)^{\frac{2}{1+\alpha}}\left(A+\frac{B}{a^{3+3\alpha}}\right)^{\frac{1}{1+\alpha}}\\
&-& 3Ba^{-3\alpha}\left(\frac{1}{\pi}\right)^{\frac{2}{1+\alpha}}\left(A+\frac{B}{a^{3+3\alpha}}\right)^{\frac{-\alpha}{1+\alpha}}
\end{array}
\right.
\label{eqsHamilton}
\end{equation}

\begin{figure}[!h]
\begin{center}
\includegraphics[height=6cm,width=6cm]{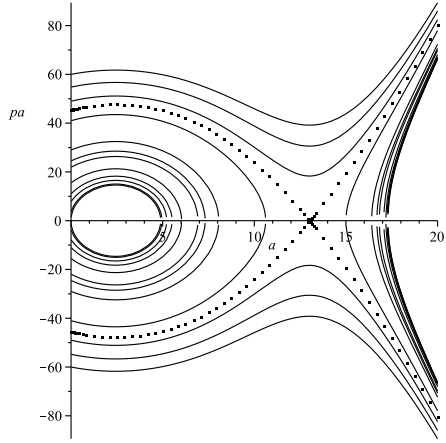}
\end{center}
\caption{Phase portrait. The curves represent Friedmann-Robertson-Walker spacetimes with $k=1$ and two fluids: a radiation perfect fluid and a generalized Chaplygin gas. 
The generalized Chaplygin gas parameters have the following values: $\alpha=0.98$, $A=0.001$ and $B=2000$.}
\label{retrato1}
\end{figure}

The general study about the possible trajectories of that system may be performed by observing the phase portrait of the model. For the present model, the phase portrait has
qualitatively the same structure for both cases: the barrier and the well followed by a barrier. As an example, we show Figure \ref{retrato1}.
In that figure, we consider the case where the potential has the shape of a well followed by a barrier, where the generalized Chaplygin gas parameters have the following values: $\alpha=0.98$, $A=0.001$ and $B=2000$. Observing Fig. \ref{retrato1}, we see that
there is a fixed point at: ($P_a=0$, $a=13.00954898$), for the surface energy $P_T=172.0907995$. That fixed point is a hyperbolic saddle. It represents the Einstein's static universe.
The dotted lines in Fig. \ref{retrato1} are called separatrices because they separate different types of solutions. For $P_a>0$ or $P_a<0$ and $P_T<172.0907995$, the solutions are bounded. It means that the universe
starts to expand from a minimum size (that size may be zero), reaches a maximum size and then contracts back to the initial size. On the other hand, for $P_a>0$ or $P_a<0$ and $P_T>172.0907995$,
the solutions are unbounded. It means that the universe starts to expand from a minimum size (that size may be zero), then it contracts for a while and finally continues to expand to an
infinity size.

\section{The quantization of the model}

In this section we are going to quantize the model by obtaining the appropriate Wheeler-DeWitt equation \cite{dewitt}. We start by introducing the wave function of the universe, 
which depends on the canonical variables $a$ and $T$,
\begin{equation}  
\Psi\, =\, \Psi(a ,T )\, .
\label{sec3eq1}
\end{equation}
Next, we promote the canonical momenta $P_{a}$ and $P_{T}$ to operators,
\begin{equation}
P_{a}\rightarrow - i\frac{\partial}{\partial a},\hspace{0.5cm} P_{T}\rightarrow - i\frac{\partial}{\partial T}.
\label{sec3eq1,5}
\end{equation}
After that, we write the Hamiltonian Eq. (\ref{sec1eq1}) as an operator ($\hat{H}$) and demand that it annihilates the wave function Eq. (\ref{sec3eq1}),
\begin{equation}
\hat{H}\Psi(a,\,T)=0.
\label{sec3eq2}
\end{equation}
Therefore, the resulting Wheeler-DeWitt equation assumes the form of the following time dependent Schr\"{o}dinger equation,
\begin{equation}
\bigg(\frac{1}{12}\frac{{\partial}^2}{\partial a^2} - 3a^2 + a^4\left(\frac{1}{\pi}\right)^{\frac{2}{(1+\alpha)}}\left(A + 
\frac{B}{a^{3+3\alpha}}\right)^{\frac{1}{(1+\alpha)}}\bigg)\Psi(a,\tau) = -i \, \frac{\partial}{\partial \tau}\Psi(a,\tau),
\label{sec3eq3}
\end{equation}
where we have imposed the reparametrization $\tau= -T$.

The operator $\hat{H}$ is self-adjoint \cite{lemos} in relation to the internal product, 
\begin{equation}
(\Psi ,\Phi ) = \int_0^{\infty} da\, \,\Psi^*(a,\tau)\, \Phi (a,\tau)\, ,
\label{sec3eq4}
\end{equation}
if the wave functions are restricted to the set of those satisfying either 
$\Psi (0,\tau )=0$ or $\Psi^{\prime}(0, \tau)=0$, where the prime $\prime$
means the partial derivative with respect to $a$. Here, we consider wave 
functions satisfying $\Psi (0,\tau )=0$ and we also 
demand that $\Psi (a,\tau )\to 0$ when $a\to \infty$.
For both effective potential cases: the barrier and the well followed by a barrier,
we are going to solve the Wheeler-DeWitt equation (\ref{sec3eq3}) numerically, using 
a finite difference procedure based on the Crank-Nicolson method \cite{crank}, 
and implemented in the program MATHEMATICA 10 \cite{cassia}. In order to evaluate
if our numerical calculations could be trusted, we computed the norm of the wavefunction
for different times, for both cases, and obtained that it was always preserved.
That criterion is normally used in numerical solutions to quantum mechanical systems \cite{gil}.
In order to solve Eq. (\ref{sec3eq3}), numerically, we have to give a initial wave function,
which fixes an energy for the radiation and the initial region where $a$ may take values.
The fact that, the initial wave function energy is associated to the radiation is due to the fact
that the time variable is related to the radiation degree of freedom. Hence, the energies of the stationary 
states are associated with the energies of the radiation fluid.

\begin{figure}[!h]
\begin{center}
\includegraphics[height=5.5cm,width=5.5cm]{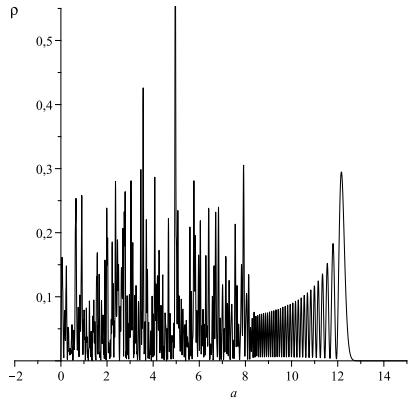}
\end{center}
\caption{$\rho=|\Psi(a,\tau_{max})|^2$, for $\alpha=0.98,\, A=0.001,\, B=0.001,\, E_m=230$, at the moment $\tau_{max}=90$, when $\Psi$ reaches the numerical infinity at $a_\infty =30$. }
\label{f1}
\end{figure}

Therefore, as initial condition, for both cases, we have chosen the following wave function,
\begin{equation}
\Psi(a,0) =\frac{8 \sqrt[4]{2}\, {E_{m}}^{3/4}\, a\, e^{-4a^2E_m}}{\sqrt[4]{\pi}}\,
\label{sec4eq1}
\end{equation}
where $E_{m}$ represents the mean kinetic energy associated to the radiation energy and $\Psi(a,0)$ is very concentrated in a region next to $a=0$. 
That initial condition is normalized in the following way: $\int_{0}^{\infty}|\Psi(a,0)|^2 da=1$. Since, for both cases, we will be interested in the tunneling probability, we will compute the outgoing wave function which has traversed the potential barrier. This implies to consider the tunneling wave function boundary condition \cite{vilenkin}. 
The portion of the wave function that tunnels the potential barrier, propagates to infinity in the positive scale factor direction, as time goes to infinity. However, we must specify a limit, in the scale factor direction, in order to perform the numerical integration of the Schr\"{o}dinger equation. Let us call that number $a_\infty$.  The behavior of these wave functions and their time evolution show that they are well defined in the whole space, even when $a$ goes to zero. As an example, we show in Figure \ref{f1} the probability density as a function of $a$, at the moment $\tau_{max}=90$, when $\Psi$ reaches the numerical infinity, at $a_\infty =30$. This example is for the case where the effective potential Eq. (\ref{sec1eq2}) is a barrier, with: $\alpha=0.98$, $A=0.001$ and $B=0.001$. For those parameters values it is possible to compute the maximum potential value ($V_{max}$): $V_{max} = 234.156142421262$, which is located at $a_{max}=12.4941638679836$. The energy for the radiation fluid is chosen
to be $E_m=230$ Eq. (\ref{sec4eq1}), which is smaller than $V_{max}$. Therefore, for the present example, the tunneling process will take place. Due to that choice of $E_m$, the wave function will reach the
potential barrier at $a_{1}=11.6321469135005$ and, after tunneling, it will leave the potential barrier at $a_{2}= 13.3004293206986$.

As we have mentioned above, we want to compute the tunneling probability for both types of potential: the barrier and the well followed by a barrier. We define the $TP$ as the probability to find the
universe to the right side of the barrier. 
Mathematically, we may compute that quantity using the following expression \cite{gil},
\begin{equation}
\label{sec5eq1}
TP = {\int_{a_{2}}^{\infty} |\Psi(a,\tau_{max})|^2 da \over 
\int_{0}^{\infty} |\Psi(a,\tau_{max})|^2 da} \, ,
\end{equation}
where, as we have mentioned above, numerically infinity has to be fixed to a maximum value of $a$ ($a_\infty$), $a_2$ is the scale factor value where, after tunneling, $\Psi$
leaves the potential barrier and $\tau_{max}$ is the moment when $\Psi$ reaches the numerical infinity $a_\infty$. The denominator of equation (\ref{sec5eq1}) is equal to $1$, since the wave function is normalized to unity.

In order to evaluate the tunneling probability as a function of $E_{m}$, $\alpha$, $A$ and $B$ one
needs to fix the values of all parameters with the exception of the one under investigation. After that, one repeats the same procedure for the other parameters.
The values of $E_{m}$ must be varied in $\Psi(a,0)$ Eq. (\ref{sec4eq1}) and the values of $\alpha$, $A$ and $B$ must be varied in the effective potential Eq. (\ref{sec1eq2}). 
For each values of $E_{m}$, $\alpha$, $A$ and $B$, the scale factor values where $\Psi$ meets the potential barrier, to the left ($a_{1}$) and to the right ($a_{2}$), are calculated. 
For all cases, we impose the condition that $E_m$ is smaller than $V_{max}$. 

\section{The Barrier Case}
\label{subbarrier}

In the present section let us study the case where the effective potential Eq. (\ref{sec1eq2}) is a barrier. We want to evaluate how $TP$ varies with $E_{m}$, $\alpha$, $A$ and $B$.

\subsection{Tunneling probability as a function of $E_m$ and $\alpha$}
\label{subsec:tpenergy}

\begin{figure}[h!]
\begin{center}
\begin{minipage}{0.45\linewidth}
\includegraphics[height=6.0cm,width=6.0cm]{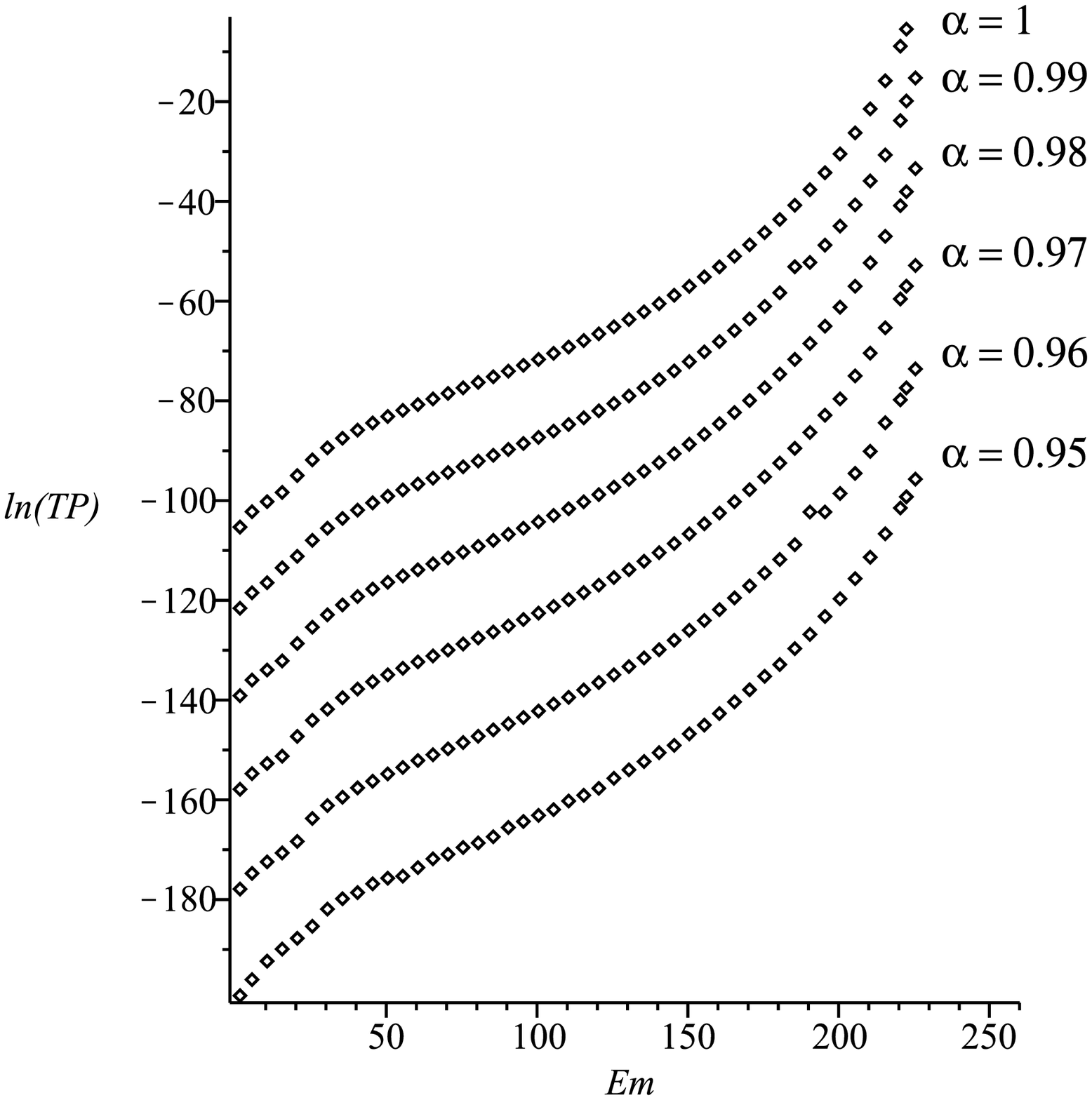}
\centerline{(a)}
\end{minipage}
\begin{minipage}{0.45\linewidth}
\includegraphics[height=6.0cm,width=6.0cm]{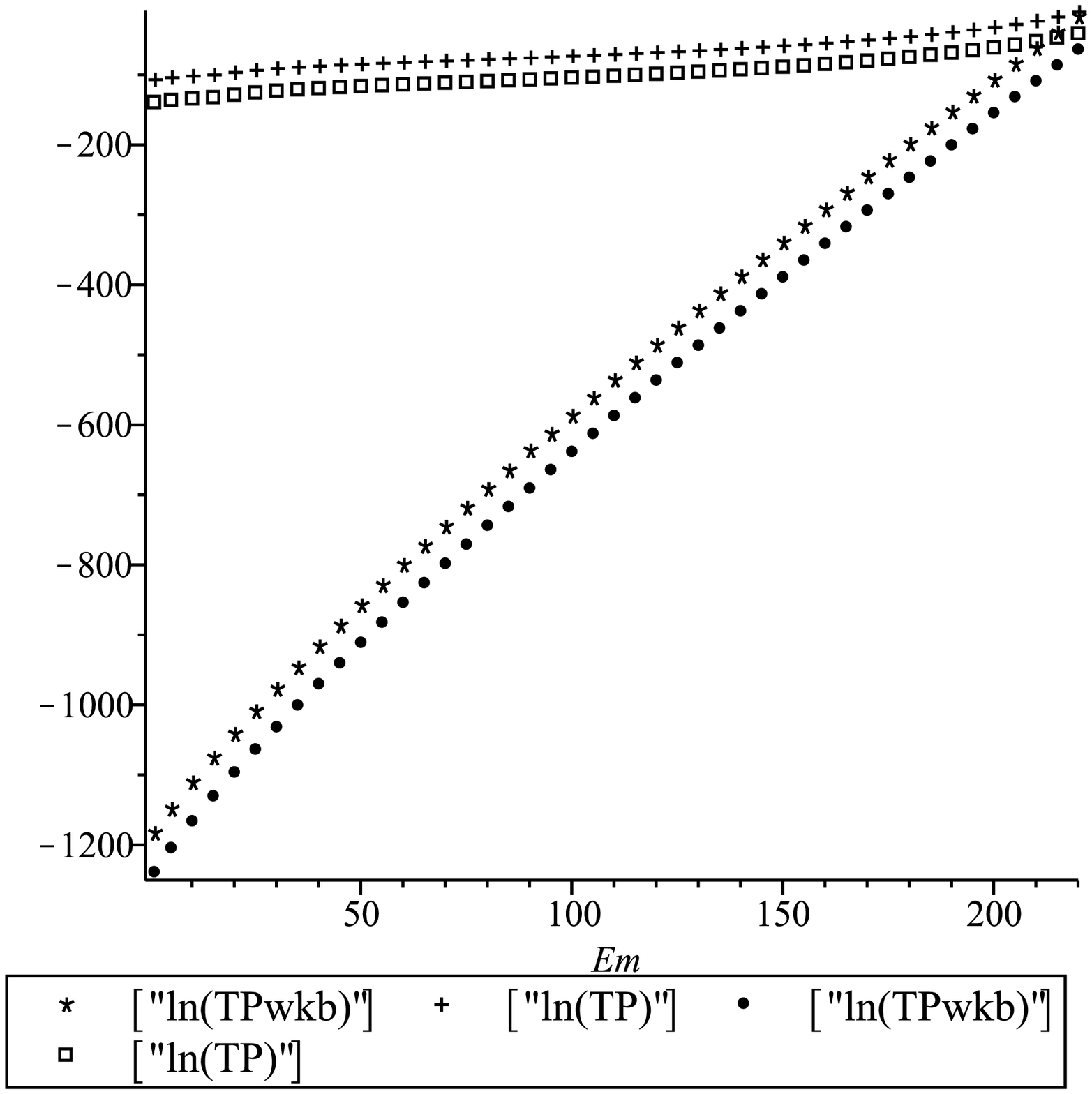}
\centerline{(b)}
\end{minipage}
\end{center}
\caption{ (a) Tunneling probability as a function of the mean energy of the initial wave packet for $\alpha = 1$, $\alpha = 0.99$, $\alpha = 0.98$, $\alpha = 0.97$, $\alpha = 0.96$, and $\alpha = 0.95$. 
(b) Comparison between $TP$ and $TP_{WKB}$, as a function of the mean energy of the initial wave packet. Here, the two curves with bigger values are for $TP$ with: $\alpha=1$ (crosses) and
$\alpha=0.98$ (boxes). On the other hand, the two curves with smaller values are for $TP_{WKB}$ with: $\alpha=1$ (asterisks) and $\alpha=0.98$ (points). All models are for the case where $A=0.001$, $B=0.001$ and $a_\infty=30$.}
\label{probabilidades-versus-Em}
\end{figure}

We start with the investigation on how $TP$ varies with $E_{m}$ and $\alpha$. Initially, we choose the values of $\alpha$, $A$ and $B$, in the effective potential Eq. (\ref{sec1eq2}), such that $V_{eff}(a)$ has a barrier shape. 
Then, we fix the values of $\alpha$, $A$ and $B$, and compute $TP$ for several
different values of $E_m$. After that, we repeat the same calculations for different values of $\alpha$. For each different value of $\alpha$, we keep $\alpha$, $A$ and $B$ fixed while we vary $E_m$. In this way, we conclude that $TP$ increases when $E_m$ and $\alpha$ increase. 
As an example, after considering many different values of those parameters, we choose the following values: $\alpha=1$, $A=0.001$ and $B=0.001$. That gives rise to a barrier type of potential, such
that its maximum value is $V_{max}=223.528202206689$. For that potential, we are considering $a_\infty=30$. After that, we compute $TP$ Eq. (\ref{sec5eq1}), for $47$ different 
values of $E_m$, all of them smaller than $V_{max}$. Those values of $E_m$ must be introduced in $\Psi(a,0)$ Eq. (\ref{sec4eq1}). From those results, we conclude that $TP$ increases when $E_m$ increases. That result agrees with the one found
in Ref. \cite{gil}, for the CG.
We repeat the same procedure, now, only modifying the value of $\alpha$. We consider the following values of $\alpha$: $0.99,\, 0.98,\, 0.97,\, 0.96$ and $0.95$. Since these values of $\alpha$ 
are very close to $1$, we use the same $47$ values of $E_m$ that we used for the potential with $\alpha=1$, for those new $\alpha$'s. In particular, the values of $V_{max}$ for those values of
$\alpha$ are, respectively: $228.753766434318$, $234.156142421262$, $239.742910169279$, $245.522039405963$ and $251.501912849465$. We notice that, for those new cases $TP$, also, increases when $E_m$
increases. Besides that, observing the $6$ curves of $TP$ versus $E_m$, we conclude that the tunneling probability increases with $\alpha$ for fixed $A$, $B$ and $E_m$.  
In Figure \ref{probabilidades-versus-Em}(a), the tunneling probabilities as functions of $E_m$, for the six values of $\alpha =1,\, 0.99,\, 0.98,\, 0.97,\, 0.96,\, 0.95$, are shown for this particular example. Due to the small values of some $TP'$s, we plot the logarithms of the $TP'$s against $E_m$. As a matter of completeness, we also compute the tunneling probability, as a function of $E_m$, using the
WKB approximation ($TP_{WKB}$) \cite{merzbacher}, for two different values of $\alpha$: $1$ and $0.98$. We use, here, the same $47$ values of $E_m$, that we have used in the determination of $TP'$s behavior.
We show our results in Figure \ref{probabilidades-versus-Em}(b). Due to the small values of some $TP_{WKB}$$'$s, we plot the logarithms of the $TP_{WKB}$$'$s against $E_m$. We can see, from that figure, that $TP_{WKB}$ increases with $E_m$ and $\alpha$, in the same way as $TP$. Here, we may see that, as was explained in Refs. \cite{gil} and \cite{gil1}, $TP$ and $TP_{WKB}$ agree only for values of $E_m'$s very close to the
top of $V_{eff}$.

Since $TP$ grows with $E_m$ and $\alpha$, it is more likely for the universe, described by 
the present model, to nucleate with the highest possible values of the radiation energy and the parameter $\alpha$.
That result, for $\alpha$, is in agreement with the approximated, analytical expression 
for $TP$ as a function of $\alpha$, obtained in Ref. \cite{paulo}, with the 
boundary conditions of the tunneling wave-function \cite{vilenkin}.
Therefore, taking in account those results, it is more likely for the universe to nucleate having as its matter content 
radiation and an ordinary Chaplygin gas ($\alpha=1$).

\subsection{Tunneling probability as a function of $A$}

\begin{figure}[h!]
\begin{center}
\includegraphics[height=5.0cm,width=5.0cm]{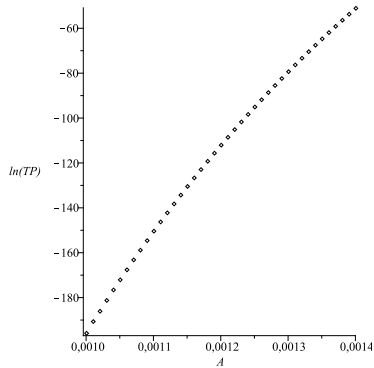}
\end{center}
\caption{Tunneling probability as a function of $A$. Here, we are considering $a_\infty=30$ and the following values of the parameters: $\alpha=0.98$, $B=0.001$ and $E_{m}= 185$.}
\label{probabilidades-versus-A}
\end{figure}

In the present subsection, we investigate how $TP$ Eq. (\ref{sec5eq1}) behaves as a function of $A$. In order to do that, initially, we fix the values of $\alpha$, $B$ and $E_m$, and compute $TP$ for several
different values of $A$. Then, we repeat the same calculations for different values of $\alpha$, still keeping $B$ and $E_m$ fixed. In this way, we conclude that $TP$ increases when $A$ increases. As an example, Figure \ref{probabilidades-versus-A} shows $TP$ as a function of $A$ for $\alpha=0.98$, $B=0.001$ and $E_{m}=185$. In this example, we used $41$ different values of $A$. Due to the small values of some $TP'$s, we plot the logarithms of the $TP'$s against $E_m$.
It is important to mention that, in Ref. \cite{gil} the authors showed that $TP$ increases with the increase of $A$, for the particular case where $\alpha=1$.

Therefore, it is more likely for the universe, described by the present 
model, to nucleate with the highest possible value of the parameter $A$.
As we have discussed in Section 2, from the energy density of the generalized 
Chaplygin gas Eq. (\ref{6}), the parameter $A$ is related, as $a$ increases, 
to a positive cosmological constant. It means that, our result indicates that
the universe should nucleate with the highest possible value of the parameter that, 
at a later time, will generate the cosmological constant.
That result is in agreement with the approximated, analytical expression 
for $TP$ as a function of $A$, obtained in Ref. \cite{paulo}, with the 
boundary conditions of the tunneling wave-function \cite{vilenkin}. It is, also,
in agreement with the result of Ref. \cite{gil1}, for a positive cosmological constant.

\subsection{Tunneling probability as a function of $B$}

\begin{figure}[h!]
\begin{center}
\includegraphics[height=5.0cm,width=5.0cm]{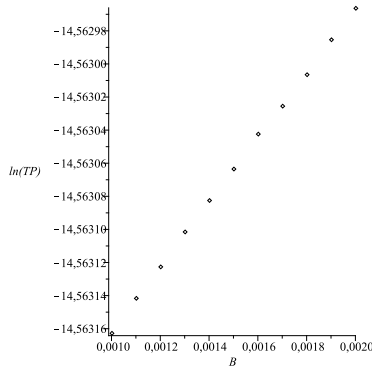}
\end{center}
\caption{Tunneling probability as a function of $B$. Here, we are considering $a_\infty=30$ and the following values of the parameters: $\alpha=0.98$, $A=0.001$ and $E_{m}= 230$.}
\label{probabilidades-versus-B}
\end{figure}

In the present subsection, we investigate how $TP$ Eq. (\ref{sec5eq1}) behaves as a function of $B$. In order to do that, initially, we fix the values of $\alpha$, $A$ and $E_m$, and compute $TP$ for several
different values of $B$. Then, we repeat the same calculations for different values of $\alpha$, still keeping $A$ and $E_m$ fixed. In this way, we conclude that $TP$ increases when $B$ increases. As an example, Figure \ref{probabilidades-versus-B} shows $TP$ as a function of $B$ for $\alpha=0.98$, $A=0.001$ and $E_{m}=230$. In this example, we used $11$ different values of $B$.
Due to the small values of some $TP'$s, we plot the logarithms of the $TP'$s against $E_m$.
It is important to mention that, in Ref. \cite{gil} the authors showed that $TP$ increases with the increase of $B$, for the particular case where $\alpha=1$.

Therefore, it is more likely for the universe, described by the present 
model, to nucleate with the highest possible value of the parameter $B$.
As we have discussed in Section 2, from the energy density of the generalized 
Chaplygin gas Eq. (\ref{6}), for small $a$, the parameter $B$ is present in the
energy density of a dust perfect fluid. It may be interpreted as the energy 
density of that dust perfect fluid, at a certain chosen time. It means that, our 
result indicates that the universe should nucleate with the highest possible value 
for the energy density of that dust perfect fluid. That result is in agreement with 
the approximated, analytical expression for $TP$ as a function of $B$, obtained in 
Ref. \cite{paulo}, with the boundary conditions of the tunneling wave-function 
\cite{vilenkin}.

\section{The well followed by a barrier case}
\label{wellbarrier}

In the present section let us study the case where the effective potential Eq. (\ref{sec1eq2}) is a well followed by a barrier. We want to evaluate how $TP$ varies with $E_{m}$, $\alpha$, $A$ and $B$.
In order to consistently describe the evolution of the wavefunction, from the initial condition Eq. (\ref{sec4eq1}), we notice that $E_m$ must be positive. On the other hand, for the present $V_{eff}$
one may have states with negative energies, inside the well. One way to solve that problem is summing the modulus of the minimum value of the well sector of $V_{eff}$ to the effective potential.
In this way, there will be no more negative energies in the well sector of $V_{eff}$. Since we have summed a constant value to the effective potential, the resulting potential will be equivalent
to the initial one. In what follows we shall use that.

\subsection{Tunneling probability as a function of $E_m$ and $\alpha$}

\begin{figure}[h!]
\begin{center}
\begin{minipage}{0.45\linewidth}
\includegraphics[height=5.5cm,width=5.5cm]{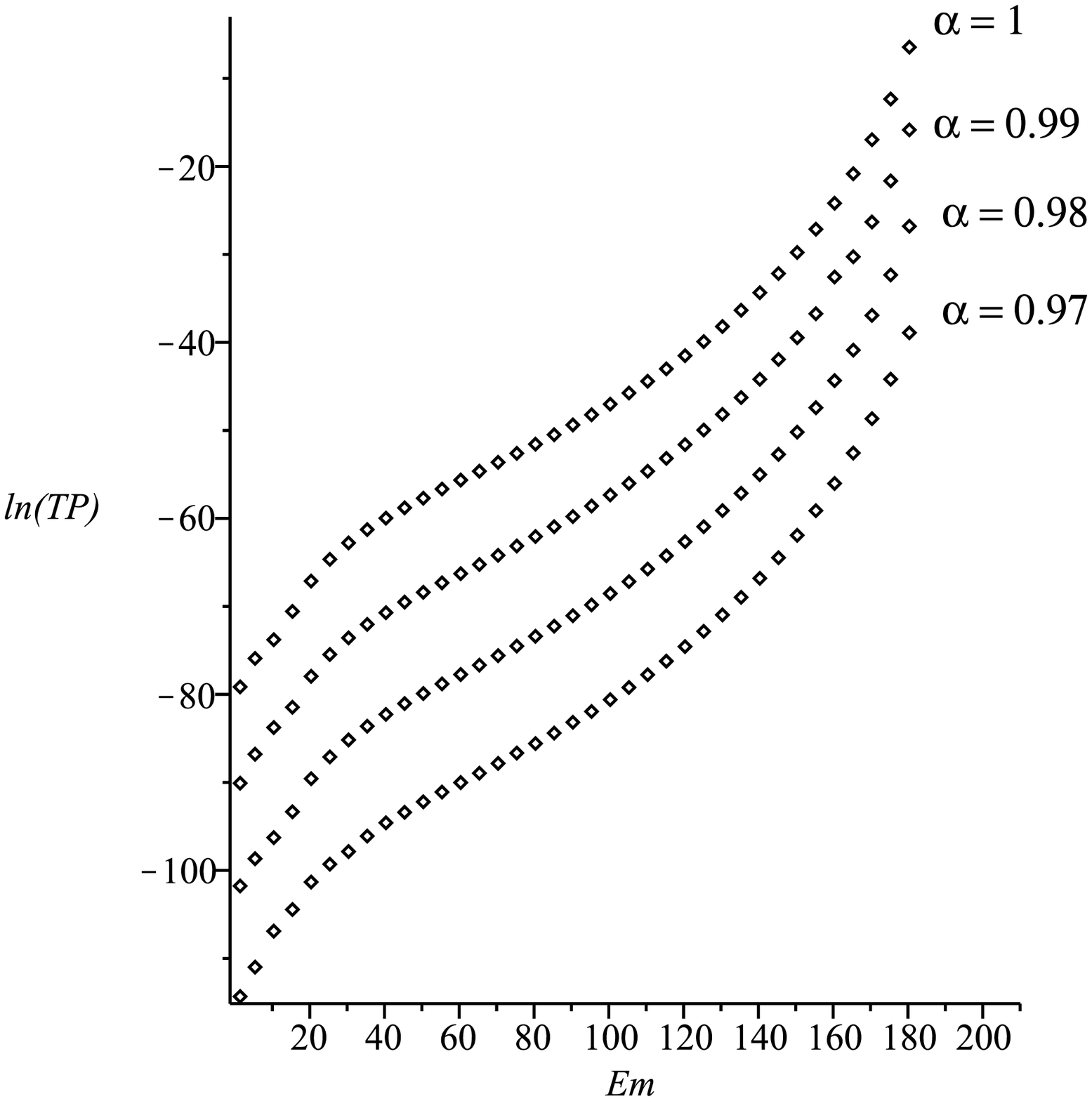}
\centerline{(a)}
\end{minipage}
\begin{minipage}{0.45\linewidth}
\includegraphics[height=6.0cm,width=6.0cm]{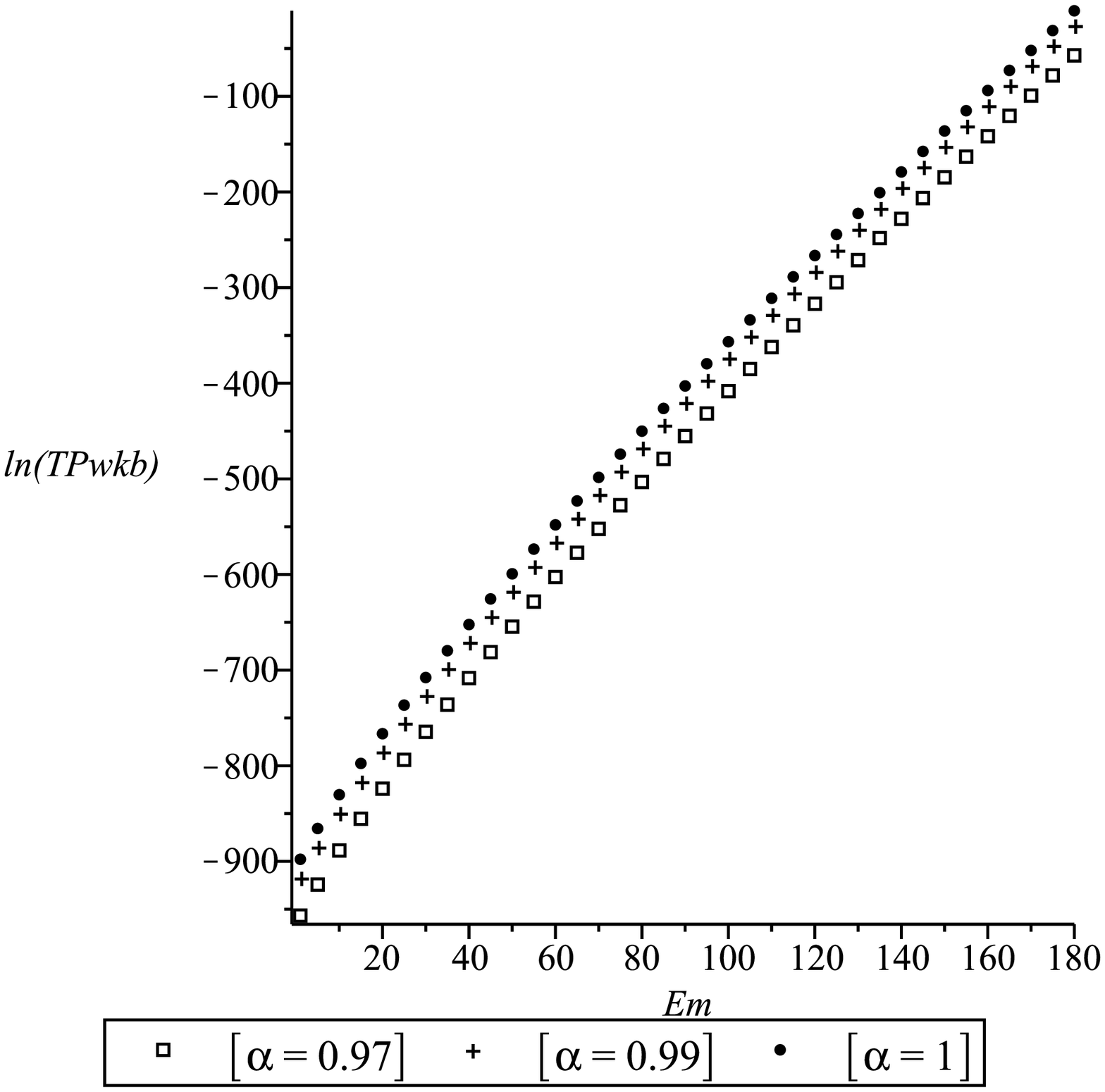}
\centerline{(b)}
\end{minipage}
\end{center}
\caption{ (a) Tunneling probability as a function of the mean energy of the initial wave packet for $\alpha = 1$, $\alpha = 0.99$, $\alpha = 0.98$ and $\alpha = 0.97$. 
(b) $TP_{WKB}$, as a function of the mean energy of the initial wave packet for three different models with: $\alpha=1$, $\alpha=0.99$ and $\alpha=0.97$. 
All models are for the case where $A=0.001$, $B=2000$ and $a_\infty=35$.}
\label{caso2probabilidades-versus-Em}
\end{figure}

In the present subsection let us study how $TP$ varies with $E_{m}$ and $\alpha$. Initially, we choose the values of $\alpha$, $A$ and $B$, in the effective potential Eq. (\ref{sec1eq2}), such that $V_{eff}(a)$ has a well followed by a barrier shape. 
Then, we fix the values of $\alpha$, $A$ and $B$, and compute $TP$ for several
different values of $E_m$. After that, we repeat the same calculations for different values of $\alpha$. For each different value of $\alpha$, we keep $\alpha$, $A$ and $B$ fixed while we vary $E_m$. In this way, we conclude that $TP$ increases when $E_m$ and $\alpha$ increase. 
As an example, after considering many different values of those parameters, we choose the following values: $\alpha=1$, $A=0.001$ and $B=2000$. That gives rise to a well followed by a barrier type of potential, such that its maximum value is $V_{max}=182.5568605$ (located at the barrier sector) and its minimum value is $V_{min}=-16.88837148$ (located at the well sector). As we have mentioned above, the modulus of $V_{min}$ shall be added to $V_{eff}$. For that potential, we are considering $a_\infty=35$. After that, we compute $TP$ Eq. (\ref{sec5eq1}), for $38$ different 
values of $E_m$, all of them smaller than $V_{max}$. Those values of $E_m$ must be introduced in $\Psi(a,0)$ Eq. (\ref{sec4eq1}). From those results, we conclude that $TP$ increases when $E_m$ increases.
We repeat the same procedure, now, only modifying the value of $\alpha$. We consider the following values of $\alpha$: $0.99,\, 0.98$ and $0.97$. Since these values of $\alpha$ 
are very close to $1$, we use the same $38$ values of $E_m$ that we used for the potential with $\alpha=1$, for those new $\alpha$$'$s. In particular, the values of $V_{max}$ for those values of
$\alpha$ are, respectively: $186.1814250$, $189.9101931$ and $193.7469173$. And the values of $V_{min}$ are, respectively: $-17.34529657$, $-17.81939352$ and $-18.31146663$. Again, we must add the modulus
of those $V_{min}$$'$s to the corresponding $V_{eff}$$'$s.
We notice that, for those new cases $TP$, also, increases when $E_m$ increases. Besides that, observing the $4$ curves of $TP$ versus $E_m$, we conclude that the tunneling probability increases with 
$\alpha$ for fixed $A$, $B$ and $E_m$. 
In Figure \ref{caso2probabilidades-versus-Em}(a), the tunneling probabilities as functions of $E_m$, for the four values of $\alpha =1,\, 0.99,\, 0.98,\, 0.97$, are shown for this particular example. 
Due to the small values of some $TP'$s, we plot the logarithms of the $TP'$s against $E_m$. We also compute the tunneling probability $TP_{WKB}$, as a function of $E_m$, for the following values of 
$\alpha$: $1$, $0.99$ and $0.97$. We use, here, the same $38$ values of $E_m$, that we have used in the determination of $TP'$s behavior. We show those results in 
Figure \ref{caso2probabilidades-versus-Em}(b). Due to the small values of some $TP_{WKB}$$'$s, we plot the logarithms of the $TP_{WKB}$$'$s against $E_m$.
We can see, from that figure, that $TP_{WKB}$ increases with $E_m$ and $\alpha$, in the same way as $TP$.

Here, as in the previous case of the barrier, $TP$ grows with $E_m$ and $\alpha$. Therefore, we may draw the same 
conclusions that we did there: it is more likely for the universe, described by the present model, to nucleate with 
the highest possible values of the radiation energy and the parameter $\alpha$.
That result, for $\alpha$, is in agreement with the approximated, analytical expression 
for $TP$ as a function of $\alpha$, obtained in Ref. \cite{paulo}, with the 
boundary conditions of the tunneling wave-function \cite{vilenkin}.
Therefore, taking in account those results, it is more likely for the universe to nucleate having as its matter content 
radiation and an ordinary Chaplygin gas ($\alpha=1$).

\subsection{Tunneling probability as a function of $A$}

\begin{figure}[h!]
\begin{center}
\includegraphics[height=5.0cm,width=5.0cm]{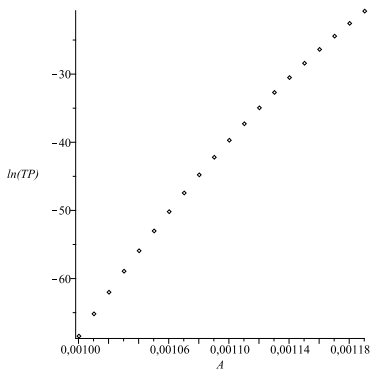}
\end{center}
\caption{Tunneling probability as a function of $A$. Here, we are considering $a_\infty=35$ and the following values of the parameters: $\alpha=0.98$, $B=2000$ and $E_{m}= 100$.}
\label{caso2probabilidades-versus-A}
\end{figure}

In the present subsection, we investigate how $TP$ Eq. (\ref{sec5eq1}) behaves as a function of $A$, for the well followed by a barrier type of potential. In order to do that, initially, we fix the values of
$\alpha$, $B$ and $E_m$, and compute $TP$ for several different values of $A$. Then, we repeat the same calculations for different values of $\alpha$, still keeping $B$ and $E_m$ fixed. In this way, we conclude that $TP$ increases when $A$ increases. As an example, Figure \ref{caso2probabilidades-versus-A} shows $TP$ as a function of $A$ for $\alpha=0.98$, $B=2000$ and $E_{m}=100$. In this example, we used $20$ different values of $A$. Due to the small values of some $TP'$s, we plot the logarithms of the $TP'$s against $E_m$.

Here, as in the previous case of the barrier, $TP$ grows with $A$. Therefore, 
we may draw the same conclusions that we did there: it is more likely for the 
universe, described by the present model, to nucleate with the highest possible 
value of the parameter $A$. Also, as in the barrier case, our result, here, indicates 
that the universe should nucleate with the highest possible value of the parameter 
that, at a later time, will generate the cosmological constant.
That result is in agreement with the approximated, analytical expression 
for $TP$ as a function of $A$, obtained in Ref. \cite{paulo}, with the 
boundary conditions of the tunneling wave-function \cite{vilenkin}. It is, also,
in agreement with the result of Ref. \cite{gil1}, for a positive cosmological constant.

\subsection{Tunneling probability as a function of $B$}

\begin{figure}[h!]
\begin{center}
\includegraphics[height=5.0cm,width=5.0cm]{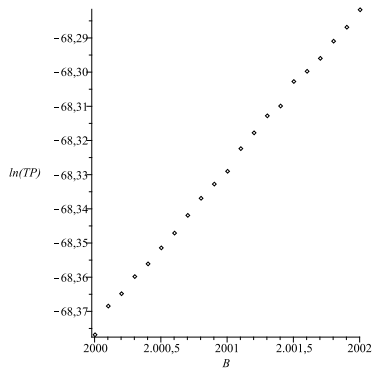}
\end{center}
\caption{Tunneling probability as a function of $B$. Here, we are considering $a_\infty=35$ and the following values of the parameters: $\alpha=0.98$, $A=0.001$ and $E_{m}= 100$.}
\label{caso2probabilidades-versus-B}
\end{figure}

In the present subsection, we investigate how $TP$ Eq. (\ref{sec5eq1}) behaves as a function of $B$, for the well followed by a barrier type of potential. In order to do that, initially, we fix the values of
$\alpha$, $A$ and $E_m$, and compute $TP$ for several different values of $B$. Then, we repeat the same calculations for different values of $\alpha$, still keeping $A$ and $E_m$ fixed. In this way, we
conclude that $TP$ increases when $B$ increases. As an example, Figure \ref{caso2probabilidades-versus-B} shows $TP$ as a function of $B$ for $\alpha=0.98$, $A=0.001$ and $E_{m}=100$. In this example, we used $21$ different values of $B$. Due to the small values of some $TP'$s, we plot the logarithms of the $TP'$s against $E_m$.

Here, as in the previous case of the barrier, $TP$ grows with $B$. Therefore, 
we may draw the same conclusions that we did there: it is more likely for the 
universe, described by the present model, to nucleate with the highest possible 
value of the parameter $B$. Also, as in the barrier case, our result, here,
indicates that the universe should nucleate with the highest possible value 
for the energy density of a dust perfect fluid. That result is in agreement with 
the approximated, analytical expression for $TP$ as a function of $B$, obtained in 
Ref. \cite{paulo}, with the boundary conditions of the tunneling wave-function 
\cite{vilenkin}.

Comparing the results of Section 4 and the present Section, we observe that $TP$ increases when $E_m$, $\alpha$, $A$ and $B$ increase, for
both shapes of the effective potential Eq. (\ref{sec1eq2}). It is an important result because it indicates that $TP$ must have that behavior
with respect to $E_m$, $\alpha$, $A$ and $B$, independently of the two effective potential shapes considered.

\section{The quantum tunneling and the initial conditions to inflation}

If the energy $E_m$ of the initial wavefunction is smaller than the maximum of $V_{eff}$, classically, the universe would
never reach an inflationary phase. On the other hand, due to the quantum tunneling effect, even for initial wavefunctions with
those values of $E_m$, the universe may undergo an inflationary phase. As it was shown in Refs. \cite{vilenkin} and \cite{vilenkin1}, it may happen even if the initial energy of the universe is nil. Therefore, for those states with $E_m < V_{eff}^{max}$, after the wavefunction has tunneled, one has the beginning of the classical universe evolution. Then, from that moment, one expects that the scale factor will follow the classical Einstein's equations. The precise classical evolution will
depend on the initial conditions. These initial conditions should be determined from the appropriate quantum states. In the
present section, we will try to find those initial conditions and determine the classical evolution.
 
For the present model, we may find the appropriate classical second order ordinary differential equation, for the scale factor, by combining the Hamilton's equations (\ref{eqsHamilton}). It is given by,
\begin{equation}
\ddot{a}+a-a^3\left(\frac{1}{\pi}\right) ^{\frac{2}{1+\alpha}}\left(A+\frac{B}{a^{3+3\alpha}}\right)^{\frac{1}{1+\alpha}}\left[\frac{2}{3}- 
\frac{B}{2a^{3+3\alpha}\left(A+\frac{B}{a^{3+3\alpha}}\right)}\right]=0,
\label{eqFriedmann}
\end{equation}
where the two dots mean the second derivative of $a$ with respect to the conformal time. Therefore, if we solve the above equation with initial conditions derived from the wavefunction that has tunneled
$V_{eff}$ (\ref{sec1eq2}), we obtain the classical evolution predicted by the quantum initial conditions. In order to do that, we need the obtain the scale factor and its time derivative from a wavefunction just 
after it has tunneled $V_{eff}$ (\ref{sec1eq2}). From a given wavefunction with a well-defined mean kinetic energy $E_m$, the expected values of the scale factor and its time derivative just after its
tunneling should be very similar to the corresponding classical values. Therefore, in order to simplify our calculations we are going to obtain the initial values of the scale factor ($a_0$) and its time
derivative ($\dot{a}_0$), using the $E_m$ of a given wavefunction and $V_{eff}$ (\ref{sec1eq2}). We do that for many different values of the parameters $\alpha$, $A$, $B$ and $E_m$, and we verify that the
resulting initial conditions, $a_0$ and $\dot{a}_0$, lead to universes that expand in an inflationary way, as it was expected due to the classical potential.
 
\begin{figure}[h!]
\includegraphics[height=5.0cm,width=5.0cm]{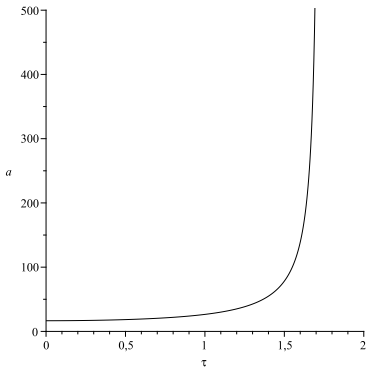}
\caption{Scale factor as a function of the conformal time for the case where $V_{eff}$ (\ref{sec1eq2}) has the barrier shape, with $\alpha=0.98$, $A=0.001$ and $B=0.001$. The mean kinetic energy is $E_m=95$,
leading to the following initial conditions: $a_{0}=16.6266317587026$ and $\dot{a}=0$.}
\label{inflacaobarreira}
\end{figure}

\begin{figure}[h!]
\includegraphics[height=5.0cm,width=5.0cm]{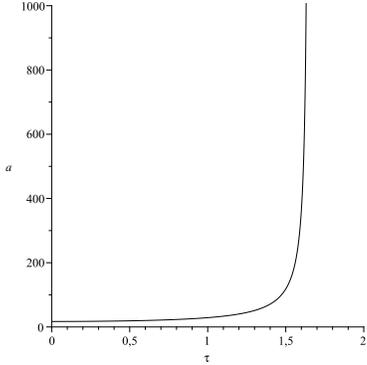}
\caption{Scale factor as a function of the conformal time for the case where $V_{eff}$ (\ref{sec1eq2}) has the well followed by a barrier shape, with $\alpha=0.99$, $A=0.001$ and $B=2000$. The mean kinetic energy is $E_m=25$, leading to the following initial conditions: $a_{0}=17.01600018$ and $\dot{a}=0$.}
\label{inflacaopocobarreira}
\end{figure}

As two examples of that result, we show Figures \ref{inflacaobarreira} and \ref{inflacaopocobarreira}. In Figure \ref{inflacaobarreira}, we have an example for the case where $V_{eff}$
(\ref{sec1eq2}) has a barrier shape, with $\alpha=0.98$, $A=0.001$ and $B=0.001$. For that case we have $E_{m}=95$, which gives the following initial conditions: $a_{0}=16.6266317587026$ and $\dot{a}_0=0$.
In Figure \ref{inflacaopocobarreira}, we have an example for the case where $V_{eff}$ (\ref{sec1eq2}) has a well followed by a barrier shape, with $\alpha=0.99$, $A=0.001$ and $B=2000$. For that case we have
$E_{m}=25$, which gives the following initial conditions: $a_{0}=17.01600018$ and $\dot{a}_0=0$. From both figures, we may see that the scale factor expands in an inflationary way.

\section{Conclusions}

In the present paper, we studied the birth of a Friedman-Robertson-Walker (FRW) universe, with positive spatial 
sections, based on GR, coupled to a radiation perfect fluid and a generalized Chaplygin gas, due to a quantum tunneling 
process. We canonically quantized the model using the Dirac's formalism for constrained systems
\cite{dirac,dirac1,dirac2}. As the result of that quantization process, we obtained a Wheeler-DeWitt equation in the
form of a time dependent Schr\"{o}dinger equation, with the effective potential $V_{eff}$ (\ref{sec1eq2}). Depending on 
the values of the GCG parameters, $V_{eff}$ may have two distinct shapes. The first one, is the shape of 
a barrier and the second one is the shape of a well followed by a barrier. Quantum mechanically, the universe may tunnel
through the potential barriers, present in both cases of $V_{eff}$. Here, we computed the tunneling probability ($TP$) for both
cases. In order to do that, we solved numerically the appropriate time dependent Schr\"{o}dinger equation and obtained the 
time evolution of an initial wave packet, for both cases. For both cases, we computed the tunneling probability, which is a function of $E_m$ and of the three parameters of the generalized Chaplygin gas: $\alpha$, $A$ and $B$.
The tunneling probabilities, for both shapes of $V_{eff}$, indicate that the universe should nucleate with the highest
possible values of $E_m$, $\alpha$, $A$ and $B$.
If we take into account that: (i) $A$ is related to the cosmological constant ($\Lambda$), for great values of the scale factor; (ii) $B$ is related to the constant present in the energy density of a dust perfect fluid, for small values of the scale factor; (iii) $\alpha$ gives the type of generalized Chaplygin gas; (iv) $E_{m}$ represents the mean kinetic energy associated to the radiation energy; then, from the tunneling probabilities, we conclude
that the universe should nucleate with the highest possible values of those quantities. In particular, for the case of $\alpha$ it means that the most probable Chaplygin gas is the ordinary one ($\alpha=1$).
The fact that $TP$ behaves, with respect to $E_m$, $\alpha$, $A$ and $B$, in the same way, for both potential shapes, is an important result.
Because it indicates that the $TP$ behavior must be independent of the two effective potential shapes.
Finally, from the appropriate tunneling wavefunction, with a suitable approximation, we obtained the values of the scale factor and its first derivative with respect to the conformal time, just after the universe has emerged from the right of the barrier. Using these initial conditions and the classical dynamical equations, we computed the classical evolution of the scale factor and showed that it leads to an universe that expand in an inflationary way.

\section*{Acknowledgments}

C. G. M. Santos thanks CNPq for her scholarship. The authors thank Paulo Vargas Moniz for discussions at an early stage of this work.


\begin{thebibliography}{}

\bibitem{dewitt} B. S. DeWitt, Phys. Rev. {\bf 160}, 1113 (1967).

\bibitem{hawking} For a complete list of references leading to that result see: S. W. Hawking and G. F. R. Ellis,
The large scale structure of space-time, (Cambridge University Press, Cambridge, 1973).

\bibitem{hawking1} J. B. Hartle and S. W. Hawking, Phys. Rev. D {\bf 28}, 2960 (1983).

\bibitem{vilenkin} A. Vilenkin, Phys. Lett. B {\bf 117}, 25 (1982).

\bibitem{vilenkin1} A. Vilenkin, Phys. Rev. D {\bf 33}, 3560 (1986).

\bibitem{riess0} A. G. Riess et al., Astron. J. {\bf 116}, 1009 (1998). 

\bibitem{perlmutter} S. Perlmutter et al., Astrophys. J. {\bf 517}, 565 (1999).

\bibitem{chaplygin} S. Chaplygin, Sci. Mem. Moscow Univ. Math. Phys. {\bf 21}, 1 (1904).

\bibitem{moschella} A. Kamenshchik, U. Moschella and V. Pasquier, Phys. Lett. {\bf B511}, 265(2001).

\bibitem{bilic} N. Bilic, G. B. Tupper and R. D. Viollier, Phys. Lett. B {\bf 535}, 17 (2002).

\bibitem{bento} M. C. Bento, O. Bertolami and A. A. Sen, Phys. Rev. D {\bf 66}, 043507 (2002).

\bibitem{paulo} M. Bouhmadi-Lopez and P.V. Moniz, Phys. Rev. D {\bf 71}, 063521 (2005).

\bibitem{gil} G.A. Monerat, G. Oliveira-Neto, E.V. Corr\^{e}a Silva, L.G. Ferreira Filho, P. Romildo Jr., 
J.C. Fabris, R. Fracalossi, S.V.B. Gon\c calves and F.G. Alvarenga. Phys. Rev. D {\bf 76}, 024017 (2007).

\bibitem{pedram0} P. Pedram and S. Jalalzadeh, Phys. Lett. B {\bf 659}, 6 (2008).

\bibitem{pedram} P. Pedram, S. Jalalzadeh. Gen. Rel. Grav. {\bf 42}, 745 (2010).

\bibitem{majumder} B. Majumder, Phys. Lett. B {\bf 697}, 101 (2011).

\bibitem{pedram1} H. Ardehali and P. Pedram, Phys. Rev. D {\bf 93}, 043532 (2016).

\bibitem{pedram2} H. Shababi and P. Pedram, Int. J. Mod. Phys. D {\bf 26}, 1750081 (2017).

\bibitem{ADM} R. Arnowitt, S. Deser and C. W. Misner, Gen. Relativ. Gravit. {\bf 40}, 1997 (2008).

\bibitem{schutz} B. F. Schutz, Phys. Rev. D {\bf 2}, 2762 (1970).

\bibitem{schutz1} B. F. Schutz, Phys. Rev. D {\bf 4}, 3559 (1971).

\bibitem{nasseri} R. Mansouri and F. Nasseri, Phys. Rev. D 60, 123512 (1999).

\bibitem{dirac}  P. A. M. Dirac, Can. J. Math. {\bf 2}, 129 (1950). 

\bibitem{dirac1} P. A. M. Dirac, Proc. Roy. Soc. London A {bf 249}, 326 and 333 (1958). 

\bibitem{dirac2} P. A. M. Dirac, Phys. Rev. {\bf 114}, 924 (1959).

\bibitem{rubakov} V.G. Lapchinskii and V.A. Rubakov, Theor. Math. Phys. {\bf 33}, 1076 (1977).

\bibitem{lemos} Lemos, N. A., J. Math. Phys. {\bf 37}, 1449, (1996).

\bibitem{crank} J. Crank and P. Nicolson, Proc. Cambridge Philos. Soc.
\textbf{43}, 50 (1947).

\bibitem{cassia} C. G. M. S. Mello, {\it Uso do m\'{e}todo de diferen\c{c}as finitas no esquema Crank-Nicolson em cosmologia qu\^{a}ntica}, Master Thesis in Computational Modeling,
(Instituto Polit\'{e}cnico, Universidade do Estado do Rio de Janeiro, Rio de Janeiro), p. 63 (2018).

\bibitem{merzbacher} E. Merzbacher, {\it Quantum Mechanics}, Second Edition, Wiley (1970).

\bibitem{gil1} J. Acacio de Barros, E. V. Corr\^{e}a Silva, G. A. Monerat, G. Oliveira-Neto,
L. G. Ferreira Filho and P. Romildo Jr., Phys. Rev. D {\bf 75}, 104004 (2007).

\end{thebibliography}
\end{document}